\documentclass[10pt]{article}
\usepackage{times,amsmath}
\usepackage{nips2002e}

\newcommand{\bde}{\begin{description}}

\newcommand{\ben}{\begin{enumerate}}
\newcommand{\beq}{\begin{eqnarray}}

\newcommand{\beqn}{\begin{eqnarray*}}

\newcommand{\bn}{{\mathbf n}}

\newcommand{\bq}{{\mathbf q}}
\newcommand{\bqu}{\begin{quote}}

\newcommand{\bvb}{\begin{verbatim}}

\newcommand{\ea}{\end{eqnarray}}

\newcommand{\ede}{\end{description}}
\newcommand{\een}{\end{enumerate}}
\newcommand{\eeq}{\end{eqnarray}}
\newcommand{\eeqn}{\end{eqnarray*}}

\newcommand{\equ}{\end{quote}}

\newcommand{\evb}{\end{verbatim}}

\newcommand{\<}{\langle}

\newcommand{\pmbeg}{\begin{pmatrix}}
\newcommand{\pmend}{\end{pmatrix}}

\renewcommand{\>}{\rangle}

\title{Inference of Entropies of Discrete Random Variables with
  Unknown Cardinalities}
\author{Ilya Nemenman\\
  Kavli Institute for Theoretical Physics\\
  University of California\\
  Santa Barbara, CA 93106\\
  \texttt{nemenman@kitp.ucsb.edu} }

\begin{document}

\maketitle

\begin{abstract}
  We examine the recently introduced NSB estimator of entropies of
  severely undersampled discrete variables and devise a procedure for
  calculating the involved integrals. We discover that the output of
  the estimator has a well defined limit for large cardinalities of
  the variables being studied.  Thus one can estimate entropies with
  no a priori assumptions about these cardinalities, and a closed form
  solution for such estimates is given.
\end{abstract}

\section{Introduction}
\label{intro}

Estimation of functions of a discrete random variable with an unknown
probability distribution using independent samples of this variable
seems to be an almost trivial problem known to many yet from the high
school \cite{stat}.  However, this simplicity vanishes if one
considers an extremely undersampled regime, where $K$, the cardinality
or the alphabet size of the random variable, is much larger than $N$,
the number of independent samples of the variable. In this case the
average number of samples per possible outcome (also called {\em bin}
in this paper) is less than one, relative uncertainty in the
underlying probability distribution is large, and the usual formulas
for estimation of various statistics fail miserably.  Then one has to
use the power of Bayesian statistics to a priori constraint the set of
allowable distributions and thus decrease the posterior error.
Unfortunately, due to the usual bias--variance tradeoff, decreasing
the variance this way may lead to an increased bias, i.e., the
estimator becomes a function of the prior, rather than of the
experimental data.

The situation is particularly bad for inferring the Boltzmann--Shannon
entropy, $S$, one of the most important characteristics of a discrete
variable.  Its frequentist as well as common Bayesian estimators have
low variances, but high biases that are very difficult to calculate
(see Ref.~\cite{entropy} for a review). However, recently ideas from
Bayesian model selection \cite{schwartz,mackay,vijay,nb02} were used
by Nemenman, Shafee, and Bialek to suggest a solution to the problem
\cite{nsb}. Their method, hereafter called NSB, is robust and unbiased
even for severely undersampled problems.  We will review it and point
out that it is equivalent to finding the number of yet unseen bins
with nonzero probability given $K$, the maximum cardinality of the
variable.  While estimation of $K$ by model selection techniques will
not work, we will show that the method has a proper limit as
$K\to\infty$.  Thus one should be able to calculate entropies of
discrete random variables even {\em without knowing their
  cardinality}.

\section{Summary of the NSB method}
\label{summary}

In Bayesian statistics, one uses Bayes rule to expresses posterior
probability of a probability distribution $\bq \equiv
\{q_i\},\,i=1\dots K,$ of a discrete random variable with a help of
its a priori probability, ${\mathcal P} (\bq)$.  Thus if $n_i$
identical and independent samples from $\bq$ are observed in bin $i$,
such that $\sum_{i=1}^K n_i =N$, then the posterior, $P(\bq|\bn)$, is
\begin{equation}
  P(\bq|\bn ) = \frac{P(\bn | \bq)
    {\mathcal P}(\bq)}{P(\bn)} 
  = \frac{ \prod_{i=1}^K q_i^{n_i}{\mathcal P}(\bq)}
  {\int_0^1 d^K q   \prod_{i=1}^K q_i^{n_i} 
    {\mathcal P}(\bq)}\;.
\end{equation}

Following Ref.~\cite{nsb}, we will focus on popular Dirichlet family
of priors, indexed by a (hyper)parameter $\beta$:
\begin{equation}
  {\mathcal P}_\beta(\bq) = \frac{1}{Z(\beta)}\,
  \delta\left( 1 - \sum_{i=1}^K q_i\right)
  \prod_{i=1}^K q_i^{\beta-1}\,, \;\;\;\; 
  Z(\beta) = \frac{\Gamma^K(\beta)}{\Gamma(K\beta)}\,.
  \label{P(q)}
\end{equation}
Here $\delta$--function and $Z(\beta)$ enforce normalizations of $\bq$
and ${\mathcal P}_\beta(\bq)$ respectively, and $\Gamma$ stands for
Euler's $\Gamma$--function. These priors are common in applications
\cite{karplus} since they, as well as the data term, $P(\bn|\bq)$, are
of a multinomial structure, which is analytically tractable. For
example, in Ref.~\cite{ww} Wolpert and Wolf calculated posterior
averages, here denoted as $\langle \dots \rangle_\beta$, of many
interesting quantities, including the distribution itself,
\begin{equation}
  \langle q_i\rangle_\beta = {\frac{n_i
      +\beta}{N+\kappa}}\,,\;\;\;\; \kappa \equiv K\beta\,,
  \label{estim}
\end{equation}
and the moments of its entropy, which we will not reprint here.

As suggested by Eq.~(\ref{estim}), Dirichlet priors add extra $\beta$
sample points to each possible bin. Thus for $\beta \gg N/K$ the data
is unimportant, and $P(\bq|\bn)$ is dominated by the distributions
close to the uniform one, $\bq \approx 1/K$. The posterior mean of the
entropy is then strongly biased upwards to its maximum possible value
of $S_{\rm max}= \ln K$.\footnote{In this paper the unit of entropy is
  {\em nat}. Thus all logarithms are natural.} Similarly, for $\beta
\ll N/K$, distributions in the vicinity of the frequentist's maximum
likelihood estimate, $\bq=\bn/N$, are important, and $\<S\>_\beta$ has
a strong downward bias \cite{entropy}.

In Ref.~\cite{nsb}, Nemenman et al.\ traced this problem to the
properties of the Dirichlet family: its members encode reasonable a
priori assumptions about $\bq$, but not about $S(\bq)$. Indeed, it
turns out that a priori assumptions about the entropy are extremely
biased, as may be seen from its following a priori moments.
\begin{eqnarray}
  \xi(\beta)  \equiv  \< S \left|_{N=0}\right. \>_\beta  &=& 
  \psi_0(\kappa+1) 
  -\psi_0(\beta+1) \, ,
  \label{Sap}
  \\
  \sigma^2(\beta) \equiv \< (\delta S)^2 \left|_{N=0}\right.  \>_\beta
     &=& 
  \frac{\beta+1}{\kappa +
    1}\, \psi_1(\beta+1) -\psi_1(\kappa+1) \,,
  \label{dS2ap}
\end{eqnarray}
where $\psi_m(x) = (d/dx)^{m+1} \ln \Gamma(x)$ are the polygamma
functions. $\xi(\beta)$ varies smoothly from $0$ for $\beta=0$,
through $1$ for $\beta \approx 1/K$, and to $\ln K$ for $\beta \to
\infty$. $\sigma(\beta)$ scales as $1/\sqrt{K}$ for almost all $\beta$
(see Ref.~\cite{nsb} for details). This is negligibly small for large
$K$.  Thus $\bq$ that is typical in ${\mathcal P}_\beta(\bq)$ usually
has its entropy extremely close to some predetermined
$\beta$--dependent value.  It is not surprising then that this bias
persists even after $N<K$ data are collected.

The NSB method suggests that to estimate entropy with a small bias one
should not look for priors that seem reasonable on the space of $\bq$,
but rather the a priori distribution of entropy, ${\mathcal
  P}(S(\bq))$, should be flattened.  This can be done approximately by
noting that Eqs.~(\ref{Sap},~\ref{dS2ap}) ensure that, for large $K$,
${\mathcal P}(S)$ is almost a $\delta$--function. Thus a prior that
enforces integration over all non--negative values of $\beta$, which
correspond to all a priori expected entropies between $0$ and $\ln K$,
should do the job of eliminating the bias in the entropy estimation
even for $N\ll K$. While there are probably other options,
Ref.~\cite{nsb} centered on the following prior, which is a
generalization of {\em Dirichlet mixture priors} \cite{mixt} to an
{\em infinite} mixture:
\begin{equation}
  {\mathcal P} (\bq;\beta) = \frac{1}{Z}\,
  \delta\left( 1 - \sum_{i=1}^K q_i\right)
  \prod_{i=1}^K q_i^{\beta-1} \frac{d \xi(\beta)}{d\beta} 
  \,{\mathcal P}(\beta)\,.
  \label{Pflat}
\end{equation}
Here $Z$ is again the normalizing coefficient, and the term
$d\xi/d\beta$ ensures uniformity for the a priori expected entropy,
$\xi$, rather than for $\beta$. A non--constant prior on $\beta$,
${\mathcal P}(\beta)$, may be used if sufficient reasons for this
exist, but we will set it to one in all further developments.

Inference with the prior, Eq.~(\ref{Pflat}), involves additional
averaging over $\beta$ (or, equivalently, $\xi$), but is nevertheless
straightforward. The a posteriori moments of the entropy are
\begin{eqnarray}
  \widehat{S^m} &=& \frac{\int_0^{\ln K} d\xi\, 
    \rho(\xi,\bn) \langle\, S^m \rangle_{\beta(\xi)}}
  {\int_0^{\ln K} d\xi\, \rho(\xi|\bn)}\,,\;\;\;\mbox{where the posterior
    density is}
  \label{Shat}
  \\
  \rho(\xi| \bn) &=& {\mathcal P}\left(\beta\left(\xi\right)\right)
  \frac{\Gamma(\kappa(\xi))}{\Gamma(N+\kappa(\xi))}\,
  \prod_{i=1}^K \frac{\Gamma(n_i+\beta(\xi))}{\Gamma(\beta(\xi))}\,.
  \label{rho}
\end{eqnarray}

Nemenman et al.\ explain why this method should work using the theory
of Bayesian model selection \cite{schwartz,mackay,vijay,nb02}. All
possible probability distributions, even those that fit the data
extremely badly, should be included in the posterior averaging. For
models with a larger volume in $\bq$ space, the number of such bad
$\bq$'s is greater, thus the probability of the model decreases.
Correspondingly, such contributions from the phase space factors are
usually termed {\em Occam razor} because they automatically
discriminate against bigger, more complex models. If the maximum
likelihood solution of a complex model explains the data better than
that of a simpler one,\footnote{This is usually achieved by requiring
  that models are nested, that is, all $\bq$'s possible in the simpler
  model are possible in the complex one, but not vice versa.} then the
total probability, a certain combination of the maximum likelihood and
the Occam factors, has a maximum for some non--trivial model, and the
sharpness of the maximum grows with $N$. In other words, the data
selects a model which is simple, yet explains it well.

In the case of Eq.~(\ref{Pflat}), we can view different values of
$\beta$ as different models. The smaller $\beta$ is, the closer it
brings us to the frequentist's maximum likelihood solution, so the
data gets explained better. However, as there is less smoothing
[cf.~Eq.~(\ref{estim})], smaller $\beta$ results in the larger phase
space. Thus, according to Ref.~\cite{nsb}, one may expect that the
integrals in Eq.~(\ref{Shat}) will be dominated by some $\beta^*$,
appropriate smoothing will be sharply selected, and $\widehat{\cdots}
\approx \<\cdots\>_{\beta^*}$. In the current paper we will
investigate whether a maximum of the integrand in Eq.~(\ref{Shat}),
indeed, exists and will study its properties. The results of the
analysis will lead us to an extension and a simplification of the NSB
method.

\section{Calculation of the NSB integrals}
\label{props}

We will calculate integrals in Eq.~(\ref{Shat}) using the saddle point
method. Since the moments of $S$ do not have $N$ dependence, when $N$
is large only the $\Gamma$--terms in $\rho$ are important for
estimating the position of the saddle and the curvature around it.  We
write
\begin{eqnarray}
  \rho(\xi|\bn) &=& {\mathcal P}(\beta(\xi)) 
  \exp \left[ -{\mathcal L}(\bn, \beta, K) \right]\,,\\
  {\mathcal L}(\bn, \beta, K) &=&  - \sum_i \ln \Gamma(\beta +n_i) 
  +K \ln \Gamma(\beta) -\ln \Gamma(\kappa) + \ln \Gamma(\kappa +N)\,.
  \label{lagrangian}
\end{eqnarray}
Then the saddle point (equivalently, the maximum likelihood) value,
$\kappa^* = K \beta^*$, solves the following equation obtained by
differentiating Eq.~(\ref{lagrangian}).
\begin{equation}
  \frac{1}{K} \sum_i^{n_i >0} \psi_0 (n_i +\beta^*) -
  \frac{K_1}{K}\,\psi_0(\beta^*) +\psi_0(\kappa^*) - \psi_0(\kappa^* +N) = 0\,,
  \label{saddle}
\end{equation}
where we use $K_m$ to denote the number of bins that have, at least,
$m$ counts. Note that $N>K_1>K_2>\dots$.

We notice that if $K\gg N$, and if there are at least a few bins that
have more that one datum in them, i.e., $K_1<N$, then the distribution
the data is taken from is highly non--uniform. Thus the entropy should
be much smaller than its maximum value of $S_{\rm max}$.  Since for
any $\beta =O(1)$ the entropy is extremely close to $S_{\rm max}$
(cf.~Ref.~\cite{nsb}), small entropy may be achievable only if
$\beta^* \to 0$ as $K \to \infty$.  Thus we will look for
\begin{equation}
  \kappa^* = \kappa_0 +\frac{1}{K} \kappa_1 + \frac{1}{K^2} \kappa_2 +
  \dots\,, 
  \label{expansion}
\end{equation}
where none of $\kappa_j$ depends on $K$. Plugging
Eq.~(\ref{expansion}) into Eq.~(\ref{saddle}), after a little algebra
we get the first few terms in the expansion of $\kappa^*$:
\begin{eqnarray}
  \kappa_1 &=& \sum_i^{n_i > 1} \frac{\psi_0(n_i) - \psi_0(1)}
  {K_1/\kappa_0^2 - \psi_1(\kappa_0) +\psi_1(\kappa_0 +N)}\,,\\
  \kappa_2 &=& \frac{\left[ \frac{K_1}{\kappa_0^3} +\frac{\psi_2(\kappa_0) - 
        \psi_2(\kappa_0 +N)}{2}\right] \kappa_1^2 + 
    \sum_i^{n_i>1}\kappa_0 \left[\psi_1(n_i) -\psi_1(1)\right]}
  {K_1/\kappa_0^2 - \psi_1(\kappa_0) +\psi_1(\kappa_0 +N) }\,,
\end{eqnarray}
and the zeroth order term solves the following algebraic equation
\begin{equation}
  \frac{K_1}{\kappa_0} = \psi_0(\kappa_0 +N ) - \psi_0(\kappa_0)\,.
  \label{zeroth}
\end{equation}
If required, more terms in the expansion can be calculated, but for
common applications $K$ is so big that none are usually needed.

We now focus on solving Eq.~(\ref{zeroth}).  For $\kappa_0 \to 0$ and
$N>0$, the r.\ h.\ s.\ of the equation is approximately $ 1/\kappa_0$
\cite{specfun}. On the other hand, for $\kappa_0 \to \infty$, it is
close to $N/\kappa_0$. Thus if $N=K_1$, that is, the number of
coincidences among different data, $\Delta \equiv N-K_1$, is zero,
then the l.\ h.\ s.\ always majorates the r.\ h.\ s., and the equation
has no solution.  If there are coincidences, a unique solution exists,
and the smaller $\Delta$ is, the bigger $\kappa_0$ is. Thus we may
want to search for $\kappa_0 \sim 1/\Delta + O (\Delta^0)$.

Now it is useful to introduce the following notation:
\begin{equation}
  f_N(j) \equiv \sum_{m=0}^{N-1} \frac{m^j}{N^{j+1}}\,,
\end{equation}
where each of $f_N$'s scales as $N^0$. Using standard results for
polygamma functions \cite{specfun}, we rewrite Eq.~(\ref{zeroth}) as
\begin{equation}
  \frac{1-\delta}{\kappa_0/N} = \sum_{j=0}^{\infty} (-1)^j 
  \frac{f_N(j)}{(\kappa_0/N)^j}\,.
  \label{anotherform}
\end{equation}
Here we introduced the relative number of coincidences, $\delta \equiv
\Delta/N$. Combined with the previous observation,
Eq.~(\ref{anotherform}) suggests that we look for $\kappa_0$ of the
form
\begin{equation}
  \kappa_0 = N \left( \frac{b_{-1}} 
    {\delta} +b_0
    +b_1\delta+\dots\right)\,,
  \label{bexp}
\end{equation}
where each of $b_j$'s is independent of $\delta$ and scales as $N^0$.

Substituting this expansion for $\kappa_0$ into
Eq.~(\ref{anotherform}), we see that it is self--consistent, and
\begin{eqnarray}
  b_{-1} &=& f_N(1) = \frac{N-1}{2N}\,,\\
  b_0 &=& -\frac{f_N(2)}{f_N(1)}= \frac{-2N + 1}{3N}\,,\\
  b_1 &=&  - \frac{f_N^2(2)}{f_N^3(1)}
  +\frac{f_N(3)}{f_N^2(1)}=\frac{N^2 - N -2 }{9(N^2 - N)}\,.
\end{eqnarray}
Again, more terms can be calculated if needed.

This expresses the saddle point value $\beta^*$ (or $\kappa^*$, or
$\xi^*$) as a power series in $1/K$ and $\delta$. In order to complete
the evaluation of integrals in Eq.~(\ref{Shat}), we now need to
calculate the curvature at this saddle point. Simple algebra results
in
\begin{equation}
  \left.\frac{\partial^2 {\mathcal L}}{\partial
      \xi^2}\right|_{\xi(\beta^*)} = 
  \left[\frac{\partial^2 {\mathcal L}}{\partial
      \beta^2}\frac{1}{(d\xi/d\beta)^2}\right]_{\beta^*} = \Delta +
  NO(\delta^2)\,.
  \label{curv}
\end{equation}
Notice that the curvature {\em does not} scale as a power of $N$ as
was suggested in Ref.~\cite{nsb}. Our uncertainty in the value of
$\xi^*$ is determined to the first order only by coincidences. One can
understand this by considering a very large $K$ with most of the bins
having negligible probabilities. Then counts of $n_i=1$ are not
informative for entropy estimation, as they can correspond to massive
bins, as well as to some random bins from the sea of negligible ones.
However, coinciding counts necessarily signify an important bin, which
should influence the entropy estimator. Note also that to the first
order in $1/K$ the exact positioning of coincidences does not matter:
a few coincidences in many bins or many coincidences in a single one
produce the same saddle point and the same curvature around it,
provided that $\Delta$ stays the same. While this is an artifact of
our choice of the underlying prior ${\mathcal P}_{\beta}(\bq)$ and may
change in a different realization of the NSB method, this behavior
parallels famous Ma's entropy estimator, which is also coincidence
based \cite{ma-81}.

In conclusion, if the number of coincidences, not $N$, is large, then
a proper value for $\beta$ is selected, and the variance of entropy is
small. Then the results of this section transform calculations of
complicated integrals in Eq.~(\ref{Shat}) to pure algebraic
operations. This analysis has been used to write a general purpose
software library for estimating entropies of discrete variables. The
library is available from the author.

\section{Choosing a value for $K$?}
\label{chooseK}

A question is in order now. If $N\ll K$, the regime we are mostly
interested in, then the number of extra counts in occupied bins,
$K_1\beta$, is negligible compared to the number of extra counts in
empty bins, $(K-K_1)\beta \approx K\beta$.  Then
Eqs.~(\ref{estim},~\ref{rho}) tell us that selecting $\beta$ (that is,
integrating over it) means balancing $N$, the number of actual counts
versus $\kappa=K\beta$, the number of pseudocounts, or, equivalently,
the scaled number of unoccupied bins. Why do we vary the pseudocounts
by varying $\beta$? Can we instead use Bayesian model selection
methods to set $K$? Indeed, not having a good handle on the value of
$K$ is usually one of the main reasons why entropy estimation is
difficult. Can we circumvent this problem?

To answer this, note that smaller $K$ leads to a higher maximum
likelihood value since the total number of pseudocounts is less.
Unfortunately, smaller $K$ also means smaller volume in the
distribution space since there are fewer bins, fewer degrees of
freedom, available. As a result, Bayesian averaging over $K$ will be
trivial: the smallest possible number of bins, that is no empty bins,
will dominate. This is very easy to see from Eq.~(\ref{rho}): only the
first ratio of $\Gamma$--functions in the posterior density depends on
$K$, and it is maximized for $K=K_1$. Thus straight--forward selection
of the value of $K$ is not an option. However, in the next Section we
will suggest a way around this hurdle.

\section{Unknown or infinite $K$}
\label{unknownK}

When one is not sure about the value of $K$, it is usually because its
simple estimate is intolerably large.  For example, consider measuring
entropy of $\ell$--gramms in printed English \cite{sg-96} using an
alphabet with 29 characters: 26 different letters, one symbol for
digits, one space, and one punctuation mark.  Then even for $\ell$ as
low as 7, a naive value for $K$ is $29^7 \sim 10^{10}$. Obviously,
only a miniscule fraction of all possible $7$--gramms may ever happen,
but one does not know how many exactly.  Thus one is forced to work in
the space of full cardinality, which is ridiculously undersampled.

A remarkable property of the NSB method, as follows from the saddle
point solution in Sec.~\ref{props}, is that it works even for finite
$N$ and extremely big $K$ (provided, of course, that there are
coincidences). Moreover, if $K\to\infty$, the method simplifies since
then one should only keep the first term in the expansion,
Eq.~(\ref{expansion}). Even more interestingly, for every $\beta \gg
1/K$ the a priori distribution of entropy becomes an exact delta
function since the variance of entropy drops to zero as $1/K$, see
Eq.~(\ref{dS2ap}). Thus the NSB technique becomes more precise as $K$
increases. So the solution to the problem of unknown cardinality is to
use an upper bound estimate for $K$: it is much better to overestimate
$K$ than to underestimate it. If desired, one may even assume that
$K\to\infty$ to simplify the calculations.

It is important to understand which additional assumptions are used to
come to this conclusion. How can a few data points specify entropy of
a variable with potentially infinite cardinality? As explained in
Ref.~\cite{nsb}, a typical distribution in the Dirichlet family has a
very particular rank ordered (Zipf) plot: the number of bins with the
probability mass less than some $q$ is given by an incomplete
$B$--function, $I$,
\begin{equation}
  \nu(q) = K I(q; \beta, \kappa -\beta) \equiv K \frac{\int_0^q dx x^{\beta -1}
    (1-x)^{\kappa - \beta -1}}{B(\beta, \kappa -\beta)}
  \label{zipf}
\end{equation}
where $B$ stand for the usual complete $B$--function. NSB fits for a
proper value of $\beta$ (and $\kappa = K\beta$) using bins with
coincidences, the head of the rank ordered plot. But knowing $\beta$
immediately defines the tails, where no data has been observed yet,
and the entropy can be calculated. Thus if the Zipf plot for the
distribution being studied has a substantially longer tail than
allowed by Eq.~(\ref{zipf}), then one should suspect the results of
the method. For example, NSB will produce wrong estimates for a
distribution with $q_1=0.5$, $q_2,\dots q_K = 0.5/(K-1)$, and
$K\to\infty$.

With this caution in mind, we may now try to calculate the estimates
of the entropy and its variance for extremely large $K$. We want them
to be valid even if the saddle point analysis of Sec.~\ref{props}
fails because $\Delta$ is not large enough. In this case $\beta^* \to
0$, but $\kappa^* = K \beta^*$ is some ordinary number. The range of
entropies now is $0 \le S \le \ln K \to \infty$, so the prior on $S$
produced by ${\mathcal P} (\bq;\beta)$ is (almost) uniform over a
semi--infinite range and thus is non--normalizable.  Similarly, there
is a problem normalizing ${\mathcal P}_{\beta}(\bq)$,
Eq.~(\ref{P(q)}). However, as is common in Bayesian statistics, these
problems can be easily removed by an appropriate limiting procedure,
and we will not pay attention to them in the future.

When doing integrals in Eq.~(\ref{Shat}), we need to find out how
$\<S(\bn)\>_{\beta}$ depends on $\xi(\beta)$. In the vicinity of the
maximum of $\rho$, using the formula for $\<S(\bn)\>_{\beta}$ from
Ref.~\cite{ww} we get
\begin{multline}
  \left[\<S(\bn)\>_{\kappa} - \xi(\beta)\right]\Big|_{\kappa
      \approx \kappa^*}\\=
    \frac{NK_1 -N}{(N+\kappa)\kappa}
    - \sum_i^{n_i>1} \frac{n_i \psi_0(n_i) -n_i \psi_0(1)}{N+\kappa} +
    O\left(\frac{1}{K}\right) = O(\delta,\frac{1}{K})\,.
\end{multline}
The expression for the second moment is similar, but complicated
enough so that we chose not to write it here . The main point is that
for $K\to\infty$, $\delta =\Delta/N \to 0$, and $\kappa$ in the
vicinity of $\kappa^*$, the posterior averages of the entropy and its
square are almost indistinguishable from $\xi$ and $\xi^2$, the a
priori averages. Since now we are interested in small $\Delta$
(otherwise we can use the saddle point analysis), we will use $\xi^m$
instead of $\<S^m\>_{\beta}$ in Eq.~(\ref{Shat}). The error of such
approximation is $O\left(\delta,\frac{1}{K}\right) =
O\left(\frac{1}{N},\frac{1}{K}\right)$.

Now we need to slightly transform the Lagrangian,
Eq.~(\ref{lagrangian}). First, we drop terms that do not depend on
$\kappa$ since they appear in the numerator and denominator of
Eq.~(\ref{Shat}) and thus cancel. Second, we expand around $1/K =0$.
This gives
\begin{equation}
  {\mathcal L}(\bn, \kappa, K) =  - \sum_i^{n_i>1} \ln \Gamma(n_i) 
  -K_1 \ln \kappa -\ln \Gamma(\kappa) + \ln \Gamma(\kappa +N) +
  O(\frac{1}{K})\,.
  \label{lagr_noK}
\end{equation}
We note that $\kappa$ is large in the vicinity of the saddle if
$\delta$ is small and $N$ is large, cf.~Eq.~(\ref{bexp}). Thus, by
definition of $\psi$--functions, $\ln \Gamma(\kappa +N)-\ln
\Gamma(\kappa) \approx N\psi_0(\kappa) + N^2 \psi_1(\kappa)/2$.
Further, $\psi_0(\kappa) \approx \ln \kappa$, and $\psi_1(\kappa)
\approx 1/\kappa$ \cite{specfun}. Finally, since $\psi_0(1) =
-C_\gamma$, where $C_\gamma$ is the Euler's constant, Eq.~(\ref{Sap})
says that $\xi -C_\gamma \approx \ln \kappa$. Combining all this, we
get
\begin{equation}
  {\mathcal L}(\bn, \kappa, K) \approx  - \sum_i^{n_i>1} \ln \Gamma(n_i) 
  +\Delta (\xi - C_\gamma) +\frac{N^2}{2} \exp(C_\gamma -\xi)\,,
\end{equation}
where the $\approx$ sign means that we are working with precision
$O\left(\frac{1}{N},\frac{1}{K}\right)$.

Now we can write:
\begin{eqnarray}
  \widehat{S} &\approx& C_\gamma 
  -\frac{\partial}{\partial \Delta} \ln \int_0^{\ln K} {\rm e}^{-{\mathcal L}}
  d\xi  \,,
  \label{Shat_as}
  \\
  \widehat{(\delta S)^2} &\approx& \left(\frac{\partial}{\partial
      \Delta}\right)^2
   \ln \int_0^{\ln K} {\rm e}^{-{\mathcal L}}
  d\xi  \,.
  \label{dShat_as}
\end{eqnarray}
The integral involved in these expressions can be easily calculated by
substituting $\exp(C_\gamma -\xi) = \tau$ and replacing the limits of
integration $1/K \exp( C_\gamma) \le\tau \le \exp( C_\gamma)$ by
$0\le\tau\le\infty$. Such replacement introduces errors of the order
$(1/K)^\Delta$ at the lower limit and $\delta^2\exp(-1/\delta^2)$ at
the upper limit. Both errors are within our approximation precision if
there is, at least, one coincidence. Thus
\begin{equation}
   \int_0^{\ln K} {\rm e}^{-{\mathcal L}}
  d\xi  \approx \Gamma(\Delta) \left(\frac{N^2}{2}\right)^{-\Delta}\,.
  \label{intval}
\end{equation}
Finally, substituting Eq.~(\ref{intval}) into
Eqs.~(\ref{Shat_as},~\ref{dShat_as}) we get for the moments of the
entropy
\begin{eqnarray}
  \widehat{S} &\approx& (C_\gamma - \ln 2) + 2 \ln N
  -\psi_0(\Delta)\,,
  \label{Shat_res}
  \\ 
  \widehat{(\delta S)^2} &\approx& \psi_1(\Delta)\,.
  \label{dShat_res}
\end{eqnarray}
These equations are valid to zeroth order in $1/K$ and $1/N$.  They
provide a simple, yet nontrivial, estimate of the entropy that can be
used even if the cardinality of the variable is unknown. Note that
Eq.~(\ref{dShat_res}) agrees with Eq.~(\ref{curv}) since, for large
$\Delta$, $\psi_1(\Delta) \approx 1/\Delta$. Interestingly,
Eqs.~(\ref{Shat_res},~\ref{dShat_res}) carry a remarkable resemblance
to Ma's method \cite{ma-81}.

\section{Conclusion}

We have further developed the NSB method for estimating entropies of
discrete random variables.  The saddle point of the posterior
integrals has been found in terms of a power series in $1/K$ and
$\delta$. It is now clear that validity of the saddle point
approximation depends not on the total number of samples, but only on
the coinciding ones. Further, we have extended the method to the case
of infinitely many or unknown number of bins and very few
coincidences.  We obtained closed form solutions for the estimates of
entropy and its variance.  Moreover, we specified an easily verifiable
condition (extremely long tails), under which the estimator is not to
be trusted. To our knowledge, this is the first estimator that can
boast all of these features simultaneously.  This brings us one more
step closer to a reliable, model independent estimation of statistics
of undersampled probability distributions.

\section*{Acknowledgments}
I thank William Bialek, the co--creator of the original NSB method,
whose thoughtful advices helped me in this work. I am also grateful to
Jonathan Miller, Naftali Tishby, and Chris Wiggins, with whom I had
many stimulating discussions. This work was supported by NSF Grant
No.\ PHY99-07949 to Kavli Institute for Theoretical Physics.

{ \bibliographystyle{abbuns}\bibliography{nsb2}}

\end{document}